\documentclass[preprint,3p,twocolumn]{elsarticle}

\usepackage{epsfig}

\usepackage{graphicx,latexsym,xcolor}

\usepackage{tikz}

\usepackage[english]{babel}
\usepackage{amsmath}
\usepackage{color}
\usepackage{epsfig}
\usepackage{graphicx}
\usepackage{bm}
\usepackage{times,amsmath,amssymb}
\usepackage{subfigure}
\usepackage{amsfonts}
\usepackage{amssymb}
\usepackage{color}
\newcommand{\tr}{\mathop{\mathrm{tr}}\limits}

\global\long\def\bra#1{\left\langle #1\right|}
\global\long\def\ket#1{\left|#1\right\rangle }
\global\long\def\braket#1#2{\left\langle #1|#2\right\rangle }
\global\long\def\ketbra#1#2{\ket{#1}\bra{#2}}

\global\long\def\al{\alpha}

\global\long\def\la{\lambda}

\global\long\def\bege{\begin{equation}}
\global\long\def\ende{\end{equation}}

\global\long\def\begal{\begin{align}}
\global\long\def\endal{\end{align}}

\begin{document}

\begin{frontmatter}


\title{Exact results for the entanglement in 1D Hubbard models with spatial constraints
}

 \author[1]{Ioannis Kleftogiannis\corref{mycorrespondingauthor}}
\cortext[mycorrespondingauthor]{Corresponding author}
\ead{ph04917@yahoo.com}
\address[1]{Physics Division, National Center for Theoretical Sciences, Hsinchu 30013, Taiwan}
\author[2]{ Ilias Amanatidis }
\address[2]{Department of Physics, Ben-Gurion University of the Negev, Beer-Sheva 84105, Israel}
\author[3,4,5]{ Vladislav Popkov }
\address[3]{Department of Physics,  University of Ljubljana,
Jadranska 19, SI-1000 Ljubljana, Slovenia}
\address[4]{Department of Physics, Bergische Universit\"at Wuppertal
42097 Wuppertal}
\address[5]{HISKP, University of Bonn,
 Nussallee 14-16, 53115 Bonn, Germany}

 \date{\today}
\begin{abstract}
We investigate the entanglement in Hubbard models of hardcore bosons in $1D$,
with an additional hardcore interaction on nearest neighbouring sites.
We derive analytical formulas for the bipartite entanglement entropy for any number of particles and system size, whose ratio determines the system filling. At the thermodynamic limit the entropy diverges logarithmically for all fillings, except for half-filling, with the universal prefactor $1/2$ due to partial permutational invariance. We show how maximal entanglement can be achieved by controlling the interaction range between the particles and the filling which determines the empty space in the system. Our results show how entangled quantum phases can be created and controlled, by imposing spatial constraints on states formed in many-body systems of strongly interacting particles.
\end{abstract}
\end{frontmatter}

\section{Introduction}
Entanglement is a key concept in understanding how quantum orders
manifest in systems with many interacting
particles\cite{kitaev1,amico,horodecki},
such as the well known example of topological order\cite{Gu,Kitaev2,Kitaev3,Levin,Li,haldane0,Varney,Isakov,AKLT}.
Essentially the degree of entanglement can be used
as a measure for the strength of the
quantum correlations in a many-body system.
In the last decades enormous advancement
has been achieved in coming up with ways
to quantify the quantum orders based on entanglement
measures. Such well known examples are the entanglement entropy\cite{kitaev1,Popkov,Hamma,wang2}
or the entanglement spectrum\cite{Li,Alba,Calabrese,Pollmann}.
These measures require splitting
the system in different partitions,
whose reduced density matrix
can be used to calculate the entanglement entropy of each respective partition.
The scaling of this entanglement entropy with
the partition size, reveals important properties of the system,
such as quantum criticality.\cite{kitaev1,Popkov,Hamma,eisert}

Due to the large complexity
of the many-body systems under investigation,
which usually contain an enormous number of particles,
exact/analytical solutions of these problems are
rarely possible and difficult to obtain.

Exact methods  (Bethe Ansatz and  quantum inverse scattering method) are limited to
 the $1D$ Fermi-Hubbard model \cite{HubbardBook} and to some extent to the  Bose-Hubbard model in the low-density regime \cite{HubbardBoseBetheAnsatz},
 while approximate and numerical approaches are used to study  various aspects of the Hubbard model and
 its extensions, see e.g. a recent review  \cite{CarmeloSacramento}.
 Among other extensions, a Hubbard model with additional integrability-breaking 
 nearest-neighbor interactions
 was studied recently, showing an intriguing new phase of a quantum disentangled liquid \cite{HubbardWithNN}.

In our paper we focus on purely nearest-neighbor interaction effects and
derive exact/analytical results
for the entanglement in 1D Hubbard models of hardcore bosons\cite{wang2,hen,Nishimoto},
with additional spatial constraints imposed by the
nearest-neighbor interactions.
Many-body/Fock states
manifest as the ground states of these
Hubbard models, as the particles organize in different
spatial configurations\cite{ioannis}.
In this paper we provide analytical solutions for the density
matrix and the entanglement entropy for superpositions of such states,
for any number of particles and system size.
We study the bipartite entanglement
and show how it varies for different system fillings.
At the thermodynamic limit we find
that the entropy diverges logarithmically for all fillings
except for half-filling, with a universal prefactor 1/2.
In addition, we show how the maximal entanglement
can achieved by varying the filling.
In overall, our results provide
a way to tune the entanglement
in Hubbard models with strong interactions,
based on the empty space in the system and the interaction range
between the particles.

\section{Model}
The states studied in this paper can
be obtained by considering the ground state of a
$1D$ Hubbard-like Hamiltonian with 
nearest-neighbor interactions \cite{ioannis}, in the limit of large  
interaction $U$, when the hopping part can be neglected, i.e.
\begin{equation}
H_U = U\sum_{i=1}^{M-1} n_{i} n_{i+1},
\label{Hamiltonian}
\end{equation}
where $n_{i}=c_{i}^{\dagger}c_{i}$ is the particle number operator, with $c_{i}^{\dagger},c_{i}$ being the  creation and annihilation operators for spin-less hardcore
bosons on site $i$.
The ground state of this system filled by $N<M/2$ hardcore bosons (hereafter also called particles)
has a large degeneracy, since every spatial configuration
of the particles respecting the hard-core restriction on sites
which are nearest-neighbours (apart from the on-site hardcore restriction),
has the lowest energy (see figure~\ref{fig_1} for an illustration).
We use the name NN for these states.
The NN states are separated from the
first excited states by an energy gap,
equal to the strength of the nearest-neighbor
interaction U.
A superposition of all possible NN states with equal
amplitudes, has the form
\begin{equation}
\ket{\Psi}\equiv \ket{M,N} = \frac{1}{\sqrt{d(M,N)}}\sum_P^{d(M,N)} \ket{1 0 1 0 1 0 0  \ldots}
\label{Psi}
\end{equation}
\begin{equation}
d(M,N) = \binom{M-N+1}{N},
\label{degeneracy}
\end{equation}
where $d(M,N)$ is the number of ways to distribute
the $N$ particles on $M$ sites, assuming at least one
hole between all the particles, due to the nearest-neighbor interaction. 
An appearance of the
binomial coefficients  Eq. \ref{degeneracy} in the sum Eq. \ref{Psi} signalizes a presence of permutation group symmetry in the problem. Quantifying the impact of the hardcore constraint on nearest-neighbor sites(spatial constraint) on the entanglement entropy of a block is one of our objectives.

Note that the states of type  Eq. (\ref{Psi}) also arise in Hubbard models with dynamic restrictions not allowing
cluster formation of the particles. For example, the unique groundstate of the following Hubbard-like Hamiltonian
\begin{equation}
\begin{aligned}
H &= U \sum_{i=1}^{M-1} n_i n_{i+1} \\
& +t \sum_{i=1}^{M-1}
[(1-n_{i-1}) c_i^{+} c_{i+1} (1-n_{i+2}) +h.c.],
\label{HamNNN}
\end{aligned}
\end{equation}
with $U>0,t>0$ contains contribution from all the NN states with different amplitudes, while at the limit
$t \rightarrow 0$, the state (\ref{Psi}) becomes one of degenerate groundstates of (\ref{HamNNN}).

\section{Bipartite entanglement}

 \begin{figure}
\begin{center}
\includegraphics[width=0.9\columnwidth,clip=true]{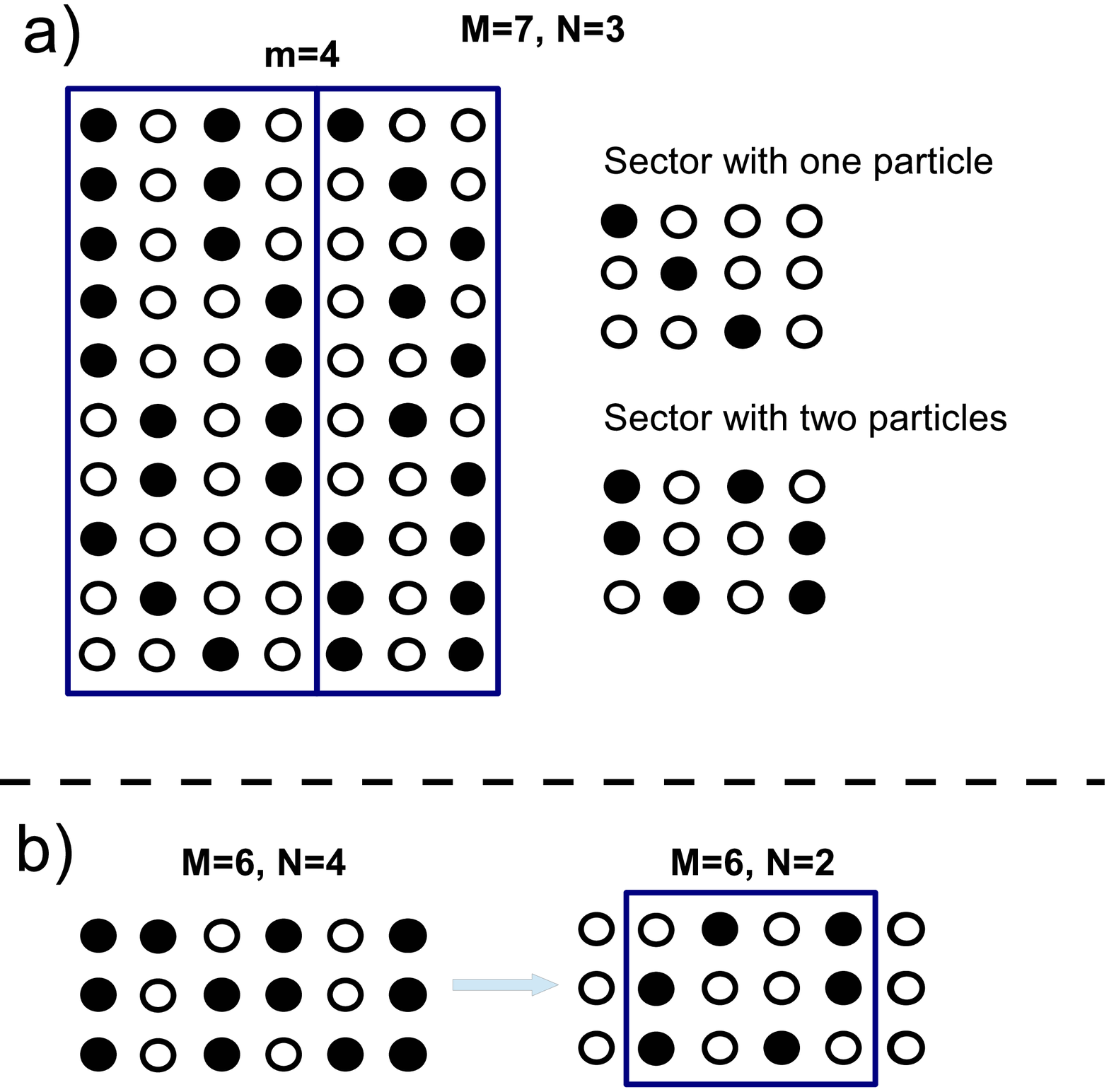}
\end{center}
\caption{a) The possible NN states with nearest-neighbor interaction
 for $N=3$ particles distributed in $M=7$ sites corresponding to filling $f=3/7$. The reduced density matrix of a partition
 containing m=4 sites, can be written in a block
 diagonal form. Each block corresponds to a sector
 according to the number of particles it contains,
 as shown on the right.
b)  NN states with $N=4$ and $M=6$ for filling $f>1/2$.
 By applying particle-hole
 exchange, the NN states transform to those corresponding to a system
 with $N=2$ and $M=6$, for $f<1/2$. This way, by ignoring the two empty edge sites
 we can calculate the entanglement entropy for $f>1/2$ by
 knowing the one for $f<1/2$.}
\label{fig_1}
\end{figure}
In this section we derive analytical results for
the reduced density matrix and the entanglement
entropy for partitioned superpositions of the NN states
described by Eq. \ref{Psi}.

Due to partial permutational symmetry enjoyed by the global pure state of the system, the
eigenvalues of the reduced density matrix can be obtained after splitting the
system of size $M$ in two parts containing  $m$ and $M-m$ sites respectively, and tracing out the
degrees of freedom of the $m$ sites. The tracing out procedure can be characterized via a permutation group analysis.

As a result of the analysis, the number of non-zero eigenvalues of the reduced density matrix grows not exponentially, but linearly with the subsystem size $m$. Origin of all the nonzero eigenvalues have been identified, as belonging to sectors with different symmetry and particle number and it was understood how to obtain them analytically, via a recursion procedure. An example of this procedure is shown schematically in figure \ref{fig_1}(a) for a small system.
The full analytic answer has been obtained for
arbitrary $N,M$. Also, the respective thermodynamic limit has been analyzed.

After tracing out the $m$ sites, and using recurrently a well known formula
\begin{align}
\binom{F+1}{N}&= \binom{F}{N-1}+\binom{F}{N},
\end{align}
after some algebra we obtain that the reduced density matrix is split into blocks
\begin{align}
\rho_{M-m}&= \sum_{k=0}^{m/2} A_{k0} \ketbra{0_k}{0_k} + A_{k1} \ketbra{1_k}{1_k}\label{BlockDensityMatrix}\\
\ket{0_k}&= \ket{M-m,N-k}\\
\ket{1_k}&= \ket{M-m-1,N-k}\otimes \ket{1,0}\\
A_{k0} d(M,N)&= \binom{m-k}{k}  \binom{M-N+1-m+k }{N-k}\\
A_{k1} d(M,N)&= \binom{m-k}{k-1}  \binom{M-N-m+k }{N-k}\\
\braket{1_k}{1_k}&= \braket{0_k}{0_k}=1.
\end{align}
Note that the property $Tr\rho_{M-m}=1$ is guaranteed by
\begin{align}
\sum_{k=0}^{m/2} (A_{k0}+A_{k1}) &=d(M,N).
\label{Completeness}
\end{align}
Now, the states $\ket{0_k},\ket{\alpha_{k'}}$ are orthogonal for $k\neq k'$ but they are not orthogonal
for $k'=k$. The overlap between $\ket{0_k},\ket{1_{k}}$ can readily be found from the combinatorial arguments to be
\begin{align}
\eta_k= \braket{0_k}{1_k}&= \sqrt{\frac {\binom{M-N-m+k }{N-k} } { \binom{M-N+1-m+k }{N-k} } }.
\end{align}
Each block with $N-k$ particles thus contains two eigenvalues $\la_k,\mu_k$, which can be found by diagonalizing the $2\times 2$ block
in (\ref{BlockDensityMatrix}).
It is then straightforward to obtain the relations
\begin{align}
 \la_k+\mu_k&=A_{k0}+A_{k1} =b\\
 \la_k \mu_k &= A_{k0}A_{k1} (1-\eta_k^2 )=c.
\end{align}
In terms of the above notations we have
\begin{align}
 \la_k &= \frac{b}{2}+ \frac{1}{2} \sqrt{b^2-4 c}
 \label{eig1}
 \\
 \mu_k &= \frac{b}{2}- \frac{1}{2} \sqrt{b^2-4 c}.
 \label{eig2}
\end{align}
The set  $\la_k$ and $\mu_k$ for all $0\leq k \leq m/2$ gives the  exact spectrum of the reduced
density matrix for arbitrary system parameters.

\subsection{Thermodynamic limit}
First, consider the limit
\begin{align}
 N\gg m\gg 1, \ \ \ N/M=f <1/2.
\end{align}
In this limit, analogically to \cite{Popkov}, and denoting
\begin{align}
 p&=\frac{f}{1-f} \\
 q&=1-p=\frac{1-2f}{1-f}\\
 n&=m-k
\end{align}
we obtain
\begin{align}
 A_{k0}& \approx \frac{1} {\sqrt{2 \pi npq}} e^{-{\frac{(k-np)^2}{2npq} }}\\
 A_{k1}& \approx \frac{1} {\sqrt{2 \pi npq}} e^{-{\frac{(k-1-np)^2}{2npq} }} \frac{N-k+1}{N-M+1 -n}
 \end{align}
valid for $npq \gg 1$. After some algebra, denoting
\begin{align}
 x&=\frac{1-k/m}{1-f}
\end{align}
we obtain
\begin{align}
 A_{k0}\equiv A_{k0}(x)&=  (1-f) \frac{1}{m(1-f)}  g(A,x)\\
 A_{k1}\equiv A_{k1}(x)&=  f \frac{1}{m(1-f)}  g\left(A,x-\frac{\kappa(f)} {m} \right)\label{Ak1}\\
 g(A,x)&= \sqrt{\frac{A}{x \pi}} e^{-A \frac {(x-1)^2}{x}}\\
 \int_0^\infty g(A,x) dx&=1\\
 \sum_k \ldots & \approx m(1-f) \int_0^\infty \ldots dx\\
A &=  m \frac{M}{M-m} \frac{1-f}{2f(1-2f)} \label{Afinitesize}.
\end{align}
Note that the last formula is valid for comparable $M \gg 1, m \gg 1, $ $m/M=const$, and
$\kappa(f)$ is of order $1$.
Finally, in the zero non-vanishing order of $1/m$, the term $\kappa(f)/m\ll 1 $ in Eq. \ref{Ak1} can be
neglected, and we obtain the final formula for the eigenvalues of the reduced density matrix (RDM) of the form

\begin{align}
 \la_{k}\equiv \la(x)&=   \frac{C_0}{m(1-f)}  g(A,x)\\
 \mu_{k}\equiv \mu(x)&=   \frac{C_1}{m(1-f)}  g(A,x)\\
C_{0},C_1&= \frac{1}{2} \pm \frac{\sqrt{1-4 f^2}}{2}.
\label{la,mu}
\end{align}
It can be proved easily that
\begin{align}
m(1-f) \int_0^\infty (\la(x)+\mu(x)) dx&=1.
\end{align}
Finally, we can find the von Neumann entropy (VNE) of the RDM, $S= -\tr \rho \log \rho$,
$\rho$ being the reduced density matrix

\begin{align}
S(f,m,M) &=-\sum_k ( \la_k \log \la_k +  \mu_k \log \mu_k)  \approx\\
- m(1-f) &\int_0^\infty \left( \la(x) \log \la(x) + \mu(x) \log \mu(x)  \right) dx.
\end{align}

\begin{figure}
\begin{center}
\includegraphics[width=0.9\columnwidth,clip=true]{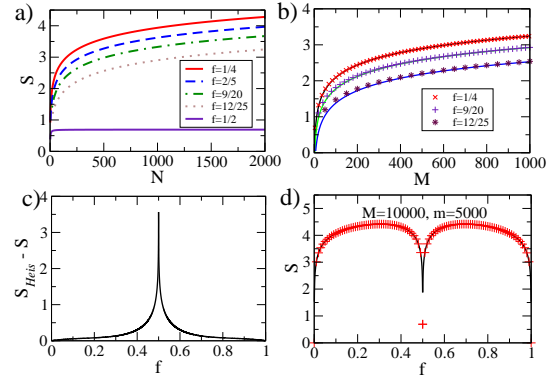}
\end{center}
\caption{a) The scaling of the bipartite entanglement entropy $S$ with
the number of particles N, for different fillings. A logarithmic divergence
at the thermodynamic limit can be observed in all cases, apart
for the half-filled case $f=1/2$.
b)Scaling of von Neumann entropy $S$ with the system size $M$. An excellent
agreement can be seen between the exact results(points) obtained
by  Eq. \ref{eig1}-\ref{eig2} and the curves obtained in the thermodynamic limit
via Eq. \ref{SVNE0}-\ref{SVNE}.
c) Comparison
with the entanglement entropy of a Heisenberg spin chain. The difference
between the entropies of the respective systems is plotted versus
the filling using Eq. \ref{SVNEHeis-SVNE} and a symmetry property
$S(f)=S( 1-f)$ established  in sec. 3.2. d) The entropy versus $f$
for a chain with $M=10000$ and $m=5000$ using Eq. \ref{SVNE}. Maximum
entanglement is obtained at $f \simeq 0.305$ and $f \simeq 0.695$. For $f=1/2$ minimum entanglement with $S=\log(2)$ is
achieved. An excellent agreement can be seen between the exact results(points) and the thermodynamic limit approximation(curve).
 }
\label{fig_2}
\end{figure}

Performing the calculations, we obtain

\begin{align}
S(f,m,M) &= Q_0(\frac{m}{M},f) + \frac{1}{2} \log m
\label{SVNE0}
\\
Q_0 &= -\sum_{\al=0,1} C_\al \log C_\al + \log\frac {(1-f)\sqrt{\pi e}} {\sqrt{A/m}}.
\label{SVNE}
\end{align}
Thus we have the same logarithmic growth of the entanglement entropy of Von Neumann, $\frac{1}{2} \log m$
as in fully permutational states of the Heisenberg ferromagnet at isotropic point, see
\cite{Popkov,spin1},
which is
apparently due to partial underlying permutational symmetry of the initial pure state. The
prefactor $1/2$ in $\frac{1}{2} \log m$ is thus simply the value of effective local spin as discussed
in \cite{spin1}.
A comparison between the exact results of Eq.  \ref{eig1}-\ref{eig2} and the thermodynamic
limit Eq. \ref{SVNE0}-\ref{SVNE} is shown on figure \ref{fig_2}b.

Note that in the form (\ref{SVNE}) an arbitrary base of logarithm can be considered. In particular, comparing (\ref{SVNE}) with the VNE computed for the ground state of the isotropic Heisenberg ferromagnet,
denoted below as $S_{Heis}$, see \cite {Popkov}, we obtain
\begin{align}
S_{Heis}- S &= \sum_{\al=0,1} C_\al \log C_\al -\frac{1}{2} \log(1-2f).
\label{SVNEHeis-SVNE}
\end{align}
As further analysis shows, $S_{Heis}> S$ for all nonzero fillings $f$.
This has the following interpretation: the ground state of the isotropic
Heisenberg ferromagnet is fully permutational invariant state with no constraints except the hard-core constraint: the minimal distance between two particles is equal to $1$: two particles can be at neighbouring sites. The wave function
Eq. \ref{Psi} has an additional constraint of a minimal distance between particles being equal to $2$. This additional constraint lowers the  symmetry of the problem, and respectively the entanglement becomes smaller.
The difference $S_{Heis}- S$, shown in  figure \ref{fig_2}c, quantifies this excess of entanglement in a state with full permutational symmetry. The
difference  $S_{Heis}- S$
increases with the filling $f$, reaching a maximum at $f=1/2$, since the effect of the additional constraint with increasing number of particles $f M=N$
becomes more and more pronounced.

Our approach of controlling
the entanglement via spatial constraints in hardcore bosonic systems,
could be applied also to other systems
that obey similar rules. One example would be spinless fermions on a chain,
since also in this case only one particle is the maximum
occupation number per site. The corresponding state described
by Eq. \ref{Psi} should contain Fock states which are antisymmetric
under exchange of two fermions, this being one of the differences
with hard-core bosons, which obey the symmetry principle instead.
Nevertheless, similar entanglement properties should be observed
to the hardcore bosonic system, as long as the fermionic system lies in the strong interaction regime, where the fermions behave as localized(point) particles.
In general, the entanglement properties of the ground state
are fully determined by the microstructure inside the Fock states in equation \ref{Psi}
along with their superposition amplitudes, irrespectively of the type of particles.

\subsection{Entanglement for $f > 1/2$}
The analysis we presented so far is valid for fillings $f < 1/2$, as we have considered a wavefunction of the form Eq. \ref{Psi}, which has at least one hole/empty site between all the particles
(minimal distance $2$ between the particles). This analysis can be easily generalized to the states for $f > 1/2$ which will contain clusters of particles and a fixed number of particle pairs. These $f > 1/2$ states
can be transformed to states with $N\rightarrow M-N$ and $M\rightarrow M-2$, i.e. to those with $f<1/2$. This can be  seen by taking the states for $f > 1/2$ (note that all  configurations
for $f>1/2$ have edge sites filled), exchanging particles with holes
and tracing out the edge sites $1$ and $M$, see figure  \ref{fig_1}b for an example. Therefore, the system of $M$ sites, $N$ particles, corresponding to $f=N/M > 1/2$ is mapped onto a system of $M'=M-2$ sites, $N'=M-N$ particles, with filling factor $f'=(M-N)/(M-2)\leq 1/2$.
Note that for odd system size $M$ and $N=(M+1)/2$, only
one NN state $\ket{1 0 1 0\ldots 1 0 1}$ contributes to the superposition Eq. \ref{Psi},
leading to $S=0$ for all $m$.
The eigenvalues of the reduced density matrix for $f=N/M >1/2$ are given by
substituting $M\rightarrow M-2$ and $N\rightarrow M-N$ in Eq. \ref{eig1}-\ref{eig2}.
In the thermodynamic limit, the configurations with fillings $f>1/2$ are mapped
onto configurations with fillings $f'=1-f$, leading to logarithmic behavior of the entropy for all $f \neq 1/2$.

\subsection{Entanglement control}
Another point of interest is the dependence of the entanglement strength
on the filling. In figure \ref{fig_2}d, we plot
$S$ versus the filling using Eq. \ref{SVNE}(curve) and compare with
the exact result using Eq. \ref{eig1}-\ref{eig2}(points).
The case $M$ odd and $N=(M+1)/2$ which gives S=0, is not present in figure  \ref{fig_2}d, since the
system size $M$ is even. The entropy is symmetric around $f=1/2$ where $S=log(2)$, due to the $f \rightarrow 1-f$ symmetry, as we have analyzed in the previous section.
The entropy obtains a maximum value
at $f \simeq 0.305$ and $f \simeq 0.695$ leading to a maximally entangled
quantum phase, irrespectively of the partitioning as long as both $fm,fM$ are large
(the asymptotic value $f \simeq 0.305884$ is obtained in the limit $m,M\rightarrow \infty$). The maximization of the entropy at this filling
is a consequence of the spatial restrictions
due to the nearest-neighbor interaction,
that impose the constraint of a minimal distance $2$ between the particles.
Changing this minimal distance by controlling the interaction range,
for example by adding a second nearest instead of nearest-neighbor interaction term
in the Hamiltonian, will lead to different fillings where the maximum entanglement occurs.
Consequently the entanglement strength in superpositions of states like
the NN ones considered in the paper,
can be tuned by the system's filling and the range of interaction between the particles.

\section{Summary and Conclusions}
We have studied analytically the entanglement
properties of NN states,
which appear as the ground states of Hubbard
chains of hardcore bosons,
with strong nearest-neighbor interactions i.e. 1D Hubbard models with spatial constraints.
We have derived exact expressions
for the entanglement entropy and the reduced
density matrix for partitioned superpositions
of the NN states. We have done that
for any number of particles and system size,
whose ratio determines the system filling.

We show that the bipartite entanglement entropy diverges
logarithmically for all fillings, apart from
half-filling, as in the critical phases of
XY spin chains.
We present a
detailed analysis of the mechanism
that creates the entanglement and make
a comparison with the entanglement of spin
chains.
The entanglement entropy
obtains a maximum value for specific fillings,
revealing a maximally entangled quantum phase.
In overall, the conditions under which this phase
occurs are determined by the spatial restrictions
imposed by the empty space in the system and 
interaction range between the particles.

In conclusion, we show analytically how
the entanglement can be tuned in
Hubbard models with strong nearest-neighbor interactions,
by controlling the empty space in the system
and the nature of the interactions between the particles,
which impose spatial restrictions on their self-organization.
We hope that our results motivate further
investigations on the mechanisms that
allow controllable entanglement
in many-body systems, to reveal novel
quantum phases of matter and help with their potential
application in the rapidly evolving field of
quantum information technology.

\section*{Acknowledgements}
IK acknowledges resources and financial support provided by
the National Center for Theoretical Sciences of R.O.C. Taiwan. IA acknowledges support from the Center for Theoretical Physics of Complex Systems in Daejeon Korea under the project IBS-R024-D1. VP acknowledges financial support from Deutsche Forschungsgemeinschaft through DFG projects KO 4771/3-1 and KL 645/20-1 and support from  ERC grant 694544  OMNES, and thanks Center for Theoretical Physics of Complex Systems in Daejeon Korea for a hospitality during his stay, during which this project has initiated.

\end{document}